\documentstyle[12pt]{article}
\begin{document}
\pagestyle{empty}
\def\singlespacing{\baselineskip=12pt}
\def\doublespacing{\baselineskip=24pt}
\doublespacing

\noindent \today \\
Submitted to Phys.\ Rev.\ E

\bigskip
\bigskip

\begin{center}
\begin{large}
{\bf ORDERING KINETICS OF CONSERVED XY MODELS} \\
\end{large}
\bigskip
\medskip
Sanjay Puri $^\dagger$, A. J. Bray and F. Rojas \\
\medskip

Theoretical Physics Group, Department of Physics and Astronomy, \\
The University, Manchester M13 9PL, UK \\

\bigskip
\bigskip

{\bf ABSTRACT}
\end{center}

\bigskip
The zero-temperature ordering kinetics of conserved XY models in spatial
dimensions $d=2$ and $3$ is studied using cell dynamical simulations.
The growth of the characteristic length scale $L(t)$ is fully consistent
with recent theoretical predictions: $L(t) \sim t^{1/4}$ for $d=2$ and
$L(t) \sim (t\ln t)^{1/4}$ for $d=3$. A gaussian closure approximation
describes the form of the structure factor rather well for $d=3$.

\bigskip
\bigskip
\bigskip
\bigskip
\bigskip

$\dagger$ Permanent address: School of Physical Sciences, Jawaharlal
Nehru University, New Delhi 110067, India.

\bigskip


\newpage
\pagestyle{plain}
\pagenumbering{arabic}

There has been considerable recent interest in the kinetics of phase
ordering in systems described by non-scalar order parameters \cite{BrayRev}.
The two fundamental questions usually addressed are (a) the extent to which
the familiar scaling phenomenology, developed in the context of scalar
systems \cite{scaling}, is applicable to non-scalar systems; and (b) if
scaling holds, the nature of the growth law for the characteristic scale
$L(t)$ that describes the coarsening dynamics, $t$ being the time.

In recent work, Bray and Rutenberg (BR) have addressed the second of
these questions using an `energy scaling' argument \cite{BR}. They obtain
predictions for the asymptotic form of $L(t)$ for all systems for which
scaling holds and for which the dynamic is purely dissipative.

The simplest non-scalar systems are XY models (or $O(2)$ systems). There
have been a number of recent simulations of the ordering kinetics of XY
systems, for both nonconserved \cite{NCOP,Rob} and conserved \cite{MG,SR}
dynamics, and some experimental studies of related liquid crystal
systems \cite{Green}. For nonconserved dynamics, the growth laws predicted
by BR are $L(t) \sim (t/\ln t)^{1/2}$ for $d=2$, and $L(t) \sim t^{1/2}$
for $d \ge 3$ (see also related work by Yurke et al.\ \cite{Pargellis}),
provided that scaling holds. For conserved dynamics, the
corresponding predictions are \cite{BR} $L(t) \sim t^{1/4}$ for $d=2$,
and $L(t) \sim (t\ln t)^{1/4}$ for $d \ge 3$.

The present work has two goals. Firstly, we present numerical results
from an extensive Cell Dynamical System (CDS) \cite{CDS} study of phase
ordering dynamics for the conserved XY model in $d=2$ and 3. In contrast
to previous numerical works, we present data for the `hardened structure
factor', enabling a clear observation of the generalized Porod tail
behavior \cite{Porod}. Our results unambiguously demonstrate dynamical
scaling for both $d=2$ and $d=3$, and are consistent with the theoretical
predictions for $L(t)$. In particular, there is clear evidence for the
predicted logarithmic correction to power-law growth for $d=3$.

The second goal of this study is to examine the utility of gaussian
closure schemes in computing the analytic form of the time-dependent
structure factor. Our analytic results using a gaussian closure scheme
are in reasonable agreement with the numerical data for $d=3$, but do not
show good agreement with the data for $d=2$. This suggests that, as is
known to be true in the nonconserved case \cite{Rob,BH}, gaussian closure
schemes for the conserved case are better for higher dimensionality.

The dynamical equation (in dimensionless form) for the ordering of the
conserved XY model has the form
\begin{equation}
\partial_t \vec{\phi} = -\nabla^2\,[\nabla^2 \vec{\phi}
                         - \partial V/\partial\vec{\phi}]\ ,
\label{MODELB}
\end{equation}
where $\vec{\phi}$ is a 2-component vector order parameter and
$V(\vec{\phi})$ has a Mexican hat form,
e.g.\ $V(\vec{\phi}) = (1-\vec{\phi}^2)^2$, with a degenerate manifold
of ground states, $|\vec{\phi}|=1$. The CDS models used in our
simulations were obtained by a conventional Euler discretization of
(\ref{MODELB}) using an isotropic discrete Laplacian. The mesh sizes of
our discretization were so large that it would be incorrect to claim that
the solution of our numerical scheme accurately shadows that of
the original partial differential equation. Thus, our models are
justifiable only as CDS models belonging to the same dynamical
universality class as the underlying partial differential equation
\cite{CDS}. Our  $d=2$ simulations were carried out on
lattices of size $256^2$ (with mesh sizes $\Delta t=0.15$ and
$\Delta x=1.7$), and the spherically averaged structure factor was
computed as an average over 80 runs with independent random initial
conditions. Our $d=3$ simulations were carried out on lattices of size
$64^3$ (with mesh sizes $\Delta t =0.1$ and $\Delta x=1.7$) and the
structure factor was calculated as an average over 50 independent runs.
The data were `hardened', i.e.\ the structure factor was computed using
fields renormalized to the length obtained from the fixed points of the
CDS iteration scheme. This procedure gives better scaling at large momenta
and elucidates the asymptotic tail behavior, which can be masked at late
times by finite defect sizes \cite{OonoPuri}. If scaling is valid,
the spherically averaged structure factor $S(k,t)$ takes the scaling form
$S(k,t) = L^d g(kL)$, where $L = L(t)$ is the characteristic scale
at time $t$, and $g(x)$ is a universal scaling function. The length $L(t)$
was defined in the usual way to be the reciprocal of the first moment,
$\langle k \rangle$, of the structure factor,
i.e.\ $L(t) = \langle k \rangle^{-1}$, where $\langle k \rangle = m_1/m_0$,
and the $m_n$ are the `moments' of the structure factor:
$m_n = \int_0^\infty dk\,k^n S(k,t)$. In this way, the length scale $L(t)$
is extracted directly from the structure factor data.

The scaled structure factor data are shown in Figures 1 ($d=3$) and 2 ($d=2$),
in (a) linear-linear and (b) log-log form. The data collapse is good for both
systems. The log-log plots reveal the large- and small-$x$ behavior of the
scaling function $g(x)$. The large-$x$ behavior follows the predicted
power-law form \cite{Porod} $g(x) \sim x^{-(d+2)}$ (`generalized Porod law'),
which is a consequence of the vortex ($d=2$) or vortex-line ($d=3$)
topological defects present in the system \cite{Humayun}.
The continuous curves, which are obtained from the approximate analytical
treatment discussed shortly, have this feature built in and thereby serve
as useful guides to the eye in the large-$x$ regime. There is a hump in
the log-log plots of Figures 1(b) and 2(b) at $x \approx 3$. This is more
clearly seen in the `Porod plot' of $g(x)\,x^{d+2}$ against $x$ (which we
will present elsewhere) and is reminiscent of the hump in the structure
factor for the conserved scalar case \cite{Hump}. For small $x$, the data
are consistent with the expected $x^4$ behavior (broken lines), which can
be derived \cite{BrayRev} using an extension of the method used to obtain
the $x^4$ behavior for scalar systems \cite{k4}.

The results for $L(t) = \langle k \rangle^{-1}$ are presented in Figure 3.
In both cases ($d=2,3$)
we attempt two different fits, a simple power-law fit $L(t) = At^x$ (Figure
3(a)), and a power-law corrected by a logarithm, $L(t) = A[t\ln (t/\tau)]^x$
(Figure 3(b)). The values of the best-fit exponent $x$ in each case, and the
timescale $\tau$ for the logarithmic fits, are shown on the Figure.
For $d=3$, the logarithmic fit is extremely good, and much better than a
simple power-law. The best-fit exponent $x=0.250$ is in perfect agreement
with the BR prediction $x=1/4$. When a simple power law is forced through
the $d=3$ data, the fit is much poorer than with the logarithmic correction,
and the value of $x$ (0.30) is unreasonably large.
By contrast, for the $d=2$ data a simple power-law gives a good fit
(much better than for $d=3$), with a best-fit exponent $x=0.247$, again
pleasingly close to the theoretical prediction $x=1/4$. The logarithmic fit
also works acceptably well for this case (as the presence
of an additional fitting parameter, i.e.\ $\tau$, would lead us to expect),
but the corresponding value of $x$ (0.21) is unreasonably small. We conclude
that the data for both dimensionalities are completely consistent with the
BR prediction.

Previous studies of the conserved XY model in $d=2$ by Mondello and
Goldenfeld \cite{MG} also found $L(t) \sim t^{0.25}$, but there was some
evidence for weak scaling violations: it was not possible to simultaneously
collapse both the position and height of the peak in $S(k,t)$. Of course,
we should point out that the results presented in \cite{MG} had a
substantially larger dynamic range (about 2.3 decades in $t$) than those
in Figure 2 (one decade in $t$). However, the data in \cite{MG} exhibit
weak scaling violations even over dynamic ranges comparable with ours.
Apart from this, we note that our statistics (average over 80
runs) are somewhat better than those of Mondello and Goldenfeld (40 runs),
and that their data were not hardened. However, is not clear to us that
either of these facts should cause such an appreciable improvement around
the peak position. In this context it is worth noting that recent studies
\cite{Rob} of the {\em nonconserved} XY model in $d=2$ found evidence
for strong scaling violations when the characteristic spacing between
vortices, $d(t) = \rho^{-1/2}$ with $\rho$ being the vortex density, was
used as the scaling length $L(t)$. It would be interesting to carry out a
similar study for the conserved case (the vortex density was not measured
in the present simulations).

Previous simulations for $d=3$ conserved XY systems \cite{SR}, on smaller
systems ($48^3$) than those studied here, were originally interpreted in
terms of power-law growth, $L(t) \sim t^{x}$. For $T=0$, the value
$x \simeq 0.29$ was obtained, close to the value $x \simeq 0.30$ obtained
here with a simple power-law fit, but the quality of the fit was not very good.
Subsequently Siegert has shown that the logarithmic form proposed by BR
(with exponent 1/4) gives a very good fit \cite{Siegert}. Some simulations
at $T>0$ were fitted to a power-law with $x \simeq 0.26$, but again the
logarithmic form (with exponent 1/4) gives a much better fit (but with a
different $\tau$ than for $T=0$) \cite{Siegert}.

We turn now to our approximate analytical treatment, based on a
`gaussian closure approximation' applied to the equation of motion.
This approach requires a straightforward modification of equivalent
treatments of nonconserved $n$-vector fields \cite{gaussian}, which in turn
generalize an earlier treatment of nonconserved scalar fields \cite{scalar}.
Here we just sketch the derivation and obtain results for the conserved
XY model ($n=2$).  A fuller treatment, with applications
to other values of $n$, will be given in a longer publication \cite{PBR}.
The starting point is the extension (\ref{MODELB}) of the Cahn-Hilliard
equation \cite{BrayRev} to vector fields. The first step is to take the
scalar product of (\ref{MODELB}), evaluated at space-time point `1', with
$\vec{\phi}(2)$, the field at space-time point `2', and average over an
ensemble of initial conditions. This gives
\begin{equation}
\partial_{t_1} C(12) = -\nabla^2\,[\nabla^2 C(12)
   -\langle \vec{\phi}(2) \cdot \partial V/\partial\vec{\phi}(1) \rangle]\ ,
\label{C}
\end{equation}
where $C(12) \equiv \langle \vec{\phi}(1) \cdot \vec{\phi}(2) \rangle$ is the
pair correlation function. In deriving (\ref{C}), we exploited the
translational invariance of the ensemble of initial conditions.
Following earlier studies of nonconserved systems \cite{gaussian}, we impose
an approximate closure of the exact equation (\ref{C}) through a gaussian
assumption for an auxiliary vector field $\vec{m}({\bf x},t)$, related to the
physical field $\vec{\phi}$ via through the equation
\begin{equation}
\nabla_m^2 \vec{\phi} = \partial V/\partial\vec{\phi}
\label{PROFILE}
\end{equation}
for the function $\vec{\phi}(\vec{m})$. Eq.\ (\ref{PROFILE}) is to be solved
with boundary conditions $\vec{\phi}(\vec{0})=0$, $\vec{\phi}(\vec{m}) \to
\hat{m}$ for $|\vec{m}| \to \infty$, where $\hat{m} = \vec{m}/|\vec{m}|$ is
a unit vector. The Laplacian operator in Eq.\ (\ref{PROFILE}) is defined
by $\nabla_m^2 \equiv \sum_{i=1}^n \partial^2/\partial m_i^2$, where the
$m_i$'s are the Cartesian components of the vector $\vec{m}$. The physical
meaning of (\ref{PROFILE}) is that the function $\vec{\phi}(\vec{m})$ gives
the structure of an equilibrium defect, with $|\vec{m}|$ regarded as the
distance from the defect core \cite{gaussian}.

The key approximation here is to treat $\vec{m}$ as a gaussian random field
in (\ref{C}). For nonconserved fields, this approximation is qualitatively
accurate, and becomes quantitatively accurate with increasing $d$
\cite{BrayRev,BH}. For conserved fields its status is less clear, but
it provides a useful starting point for the discussion of scaling functions
for conserved vector fields. (We should warn the reader that the gaussian
approximation in its present form is not even a reasonable starting point
for the conserved scalar case, which has an altogether different growth
law, i.e.\ $L(t) \sim t^{1/3}$, from the conserved vector case.) \
Using (\ref{PROFILE}) to eliminate the explicit dependence on
the potential in (\ref{C}), and exploiting the gaussian property, we obtain
a closed equation for $C(12)$. For equal-time correlations, this reads
\begin{equation}
\frac{1}{2} \frac{\partial C}{\partial t} = -\nabla^2 \left[ \nabla^2 C
   + \frac{1}{\langle m^2 \rangle}\,\gamma \frac{dC}{d\gamma} \right]\ .
\label{4}
\end{equation}
In (\ref{4}) $\langle m^2 \rangle$ is the variance of {\em one} component
$m$ of $\vec{m}$, $\gamma = \langle m(1) m(2) \rangle/\langle m^2 \rangle$
is the normalized correlator of $m$, and the function $C(\gamma)$ for general
$n$ is given by
\cite{Porod,gaussian}
\begin{equation}
C(\gamma) = \frac{n\gamma}{2\pi}\left[B\left(\frac{n+1}{2},\frac{1}{2}\right)
\right]^2\,_2F_1\left(\frac{1}{2},\frac{1}{2};\frac{n+2}{2};\gamma^2\right)\ ,
\label{C1}
\end{equation}
where $B(x,y)$ is the beta function and $_2F_1(a,b;c;z)$ is the
Hypergeometric function.

The next step is to seek a scaling solution $C=f(r/L)$, in which all the time
dependence is contained in $L = L(t)$. Requiring that all terms in (\ref{C1})
scale in the same way forces $\langle m^2 \rangle = \alpha L^2$ and
$L = (8t)^{1/4}$ (where the factor 8 is put in for convenience). Recasting
(\ref{4}) as an equation for $\gamma(x)$, where $x=r/L$ is the scaling
variable, gives
\begin{equation}
\frac{1}{f_\gamma} \nabla_x^{-2} (f_\gamma x \gamma') = \nabla_x^2 \gamma +
 \alpha\gamma + \frac{f_{\gamma\gamma}}{f_\gamma}\,(\gamma')^2\ ,
\label{gamma}
\end{equation}
which is a convenient form for numerical solution. Here the function
$f(\gamma)$ is just the right-hand side of (\ref{C1}), subscripts $\gamma$
indicate derivatives with respect to $\gamma$, primes denote derivatives
with respect to $x$, and $\nabla_x^2$ is the Laplacian with respect to $x$,
i.e.\ $\nabla_x^2 \equiv \partial^2/\partial x^2 +
[(d-1)/x]\partial/\partial x$.

In an earlier treatment of Eqs.\ (\ref{4}) and (\ref{C1}), an approximate
solution was obtained by expanding $\gamma dC/d\gamma$ in powers of $C$ and
truncating at $O(C^3)$ \cite{Rojas}. The Porod tail in the structure factor
is lost in this approximation. Despite this, the resulting real-space
correlation function was in good agreement with the data of Siegert and Rao
\cite{SR}, although this agreement may be misleading as the data in \cite{SR}
were not hardened. The present treatment is superior, as no approximations
are made beyond the initial gaussian closure. In particular, the Porod tail
is present in the solution, as is evident from Figures 1(b) and 2(b).

The left-hand side of (\ref{gamma}) can be written in integral form using the
identity \cite{Mazenko,PBR}
\begin{equation}
\nabla_x^{-2}F(x) = -\,\frac{1}{d-2}\left[\int_0^x dy\,yF(y)
\left\{\left(\frac{y}{x}\right)^{d-2}-1\right\}
+ \int_0^\infty dy\,yF(y)\right]\ ,
\label{INVLAPLACE}
\end{equation}
for $d>2$. (For $d=2$ the appropriate limit has to taken \cite{PBR}).
In the context of Eq.\ (\ref{gamma}),
$F(x) = f_\gamma x \gamma'(x) \equiv xf'(x)$ in (\ref{INVLAPLACE}).

To solve Eq.\ (\ref{gamma}) numerically requires specifying boundary
conditions at $x=0$. In this paper we specialise to the case $n=2$.
Solutions for other values of $n$ will be presented elsewhere \cite{PBR}.
For $n=2$ one can show that $\gamma(x)$ has a small-$x$ expansion of the
form
\begin{equation}
\gamma(x) = 1 - \frac{1}{2} a^2x^2 - \frac{bx^2}{\ln x} + \cdots
\label{smallx}
\end{equation}
Therefore the required boundary conditions are $\gamma(0)=1$, $\gamma'(0) = 0$.
To perform the numerical integration, however, one needs the values of
the parameter $\alpha$ in (\ref{gamma}) and of the infinite integral
$\int_0^\infty dy\,yF(y)$ in (\ref{INVLAPLACE}). Using $F(x)=xf'(x)$, this
integral can be written as $-2\int_0^\infty dy\,yf(y)$, which can be
expressed in terms of $\alpha$ and the parameters $a$ and $b$ in the
small-$x$ expansion (\ref{smallx}). Inserting (\ref{smallx}) into
(\ref{gamma}), using (\ref{INVLAPLACE}) to simplify the left-hand side,
and matching terms of $O(1)$ and $O(1/\ln x)$ gives the relations
\begin{eqnarray}
\alpha & = & da^2 \\
2\int_0^\infty dy\,y f(y) & = & (d-2)(a^2 + 2bd) \ .
\end{eqnarray}
(Again, the case $d=2$ has to be treated separately \cite{PBR}). Numerical
integration of the differential equation (\ref{gamma}) is now straightforward,
as the finite integral in (\ref{INVLAPLACE}) can be accumulated as the
integration proceeds. The two unknown parameters $a$ and $b$ are fixed by
the requirement that the large-$x$ behavior be physical, i.e.\ $f(x)$ must
vanish as $x \to \infty$. For $d=3$, these parameters take the approximate
values $a = 1.177567$, $b = -0.155217$. (For $d=2$, there are still two
parameters, but they enter in a slightly different way \cite{PBR}).

The numerical solutions for the structure factor (the Fourier transform of
the pair correlation function) are included in Figures 1 and 2 as solid
lines. Notice that there are no adjustable parameters in the fit. The fit
is surprisingly good for $d=3$, despite the fact that the growth law
$L \sim t^{1/4}$ obtained within the gaussian closure scheme does not
have the logarithmic correction predicted for the exact solution. For $d=2$,
the fit is noticeably worse. One feature of the approximate theory which is
qualitatively incorrect is the small-$k$ behavior. The log-log plots of
Figs.\ 2(b) and 3(b) show the expected $k^4$ behavior at small $k$.
This feature is missing from the gaussian closure scheme, which gives a
$k^2$ behavior at small $k$. Another point of discrepancy between the
gaussian closure theory and the numerical results is that the theory
overestimates the amplitude of the Porod tail though it replicates nicely
the qualitative features (i.e.\ valleys and humps), especially in $d=3$.
This feature is better seen in a Porod plot of theory and data, and we
will present such a plot in an extended publication.

In summary, our simulation data for the conserved XY model is fully consistent,
for both $d=2$ and $d=3$, with theoretical predictions derived from a scaling
assumption \cite{BR}. The scaling collapse of the data, using a lengthscale
$\langle k \rangle^{-1}$ extracted from the structure factor data itself, is
good. In future work it would be desirable to test scaling using a lengthscale
derived independently from, for example, the density of vortices (or vortex
lines) \cite{Rob}. An approximate calculation of the structure factor scaling
function, using a gaussian closure scheme, describes the data quite well
for $d=3$, except at small $k$. We expect this procedure to work even better
for $n>2$, where the growth law $L \sim t^{1/4}$ has no logarithmic
corrections \cite{BR}.

SP thanks the members of the Physics Department of Manchester University,
where most of this work was done, and he and AB thank the staff of the
Isaac Newton Institute, Cambridge, where the work was completed, for their
hospitality. FR thanks CONACYT (Mexico) for support.

\newpage
\begin{large}
\noindent{\bf Figure Captions}
\end{large}
\begin{enumerate}

\item Scaled structure factor for $d=3$: (a) linear-linear plot
(b) log-log plot. The data are for dimensionless times 1000, 2000, 3000 and
5000, as indicated. Details of the simulation are described in the text.
The continuous curves are the numerical Fourier transforms
of the scaling function for the pair correlation function, calculated using
the gaussian closure scheme described in the text. The dashed line in (b)
has slope 4.

\item Same as Figure 2, but for $d=2$. Here the data are for dimensionless
times 1050, 2100, 6300 and 10500.

\item Growth of the characteristic length scale $L(t)$ for the conserved XY
model, fitting to (a) $L(t) = At^x$ and (b) $L(t) = A(t\ln[t/\tau])^x$.
The best-fit values of $x$ and $\tau$ are specified on the figures.
\end{enumerate}

\end{document}